# Dynamics of bi-stripes and a colossal metal-insulator transition in the bi-layer manganite $La_{2-2x}Sr_{1+2x}Mn_2O_7$ (x~0.59)

Z. Sun[1], Q. Wang[1], A. V. Fedorov[2], H. Zheng[3], J. F. Mitchell[3], D.S. Dessau[1]

[1] *Department of Physics, University of Colorado, Boulder, CO 80309, USA*

[2] *Advanced Light Source, Lawrence Berkeley National Laboratory, Berkeley, CA 94720, USA*

[3] *Materials Science Division, Argonne National Laboratory, Argonne, IL 60439, USA*

In correlated electron materials, electrons often self-organize and form a variety of patterns with potential ordering of charges, spins, and orbitals, which are believed to be closely connected to many novel properties of these materials including superconductivity[1,2], metal-insulator transitions [3], and the CMR effect [4]. How these real-space patterns affect the conductivity and other properties of materials (which are usually described in momentum space) is one of the major challenges of modern condensed matter physics. Moreover, although the presence of static stripes is indisputable, the existence (and potential impacts) of fluctuating stripes in such compounds is a subject of great debate [5,6]. Here we present the electronic excitations of *$La_{2-2x}Sr_{1+2x}Mn_2O_7$* (*x* ~ 0.59) probed by angle-resolved photoemission (ARPES), from which we demonstrate that a novel type of ordering, termed bi-stripes [7,8], can exhibit either static or fluctuating order as a function of temperature. We found that the static bi-stripe order is especially damaging to electrical conductivity, completely localizing the electrons in the bi-stripe regions, while the fluctuating stripes can coexist with mobile carriers. This physics drives a novel phase transition with colossal conductivity changes as a function of temperature [9,10]. Our finding suggests that quantum **stripes can give rise to electronic properties significantly different from their static counterparts. Inducing transition between them can turn on remarkable electronic phenomena, enriching our understanding of correlated electron systems as well as opening a window for potential applications in electronic devices.**

Fig. 1 shows the doping phase diagram of the bi-layer manganite family $La_{2-2x}Sr_{1+2x}Mn_2O_7$ between the doping levels $x = 0.48$ and $x = 0.65$ [10]. The yellow portion covering most of the low temperature region of the phase diagram is an A-type antiferromagnetic phase (AAFM), consisting of ferromagnetically ordered metallic planes stacked in an antiferromagnetic sequence[10]. Very near the commensurate doping levels of $x = 0.50$ and $x = 0.60$, charge/orbital ordering is the stable state, as indicated by the gold and blue regions. At $x = 0.50$ the ordering is of the famous CE type first proposed by Goodenough [11]. The CE type structure bears antiferromagnetic (AFM) spin order at low temperature and paramagnetic (PM) spin order at high temperature, as illustrated in Fig. 1b and 1c [12]. This is a stripe-like state as illustrated by the diagonal grey-shaded regions. At $x = 0.60$ bi-stripe ordering, which is a variant of the CE ordering, has been proposed [7,8] and is illustrated in Fig. 1c. Compared to the CE order, the bi-stripe order has an extra row of $Mn^{4+}$ ions, accounting for the extra holes doped into the system. One of our main hypotheses is illustrated by the heavy red lines superimposed on Fig 1b: in the CE state with AFM spin order, a hopping path for electrons is drawn which follows a ferromagnetic alignment of spins and a natural overlap of orbitals – though insulating with a gap at the Fermi energy (see Fig. 2f), this hopping path nonetheless allows for Bloch-like electron states which, as evidenced by clear band dispersion in our measurements, sample multiple crystalline sites in the material. These states also may be described as coherent, though we refrain from this terminology here since a stricter definition for coherence requiring sharp Fermi Liquid-like states is sometimes applied, and is not appropriate for these relatively broad heavily-damped states. In the CE-PM state, such Bloch-like states are well sustained, though the ferromagnetic alignment is gone. On the other hand, as will be shown later, we experimentally find that in the bi-stripe phase (blue region in Fig. 1a) this hopping path is destroyed due to the static bi-stripe order (see Fig. 1d), for we have observed almost completely non-dispersive/localized electron states. A consequence of this is a colossal (∼ 5 orders of magnitude for the $x \sim 0.59$ compound) and sudden change of conductivity upon entering the bi-stripe phase from the AAFM phase below [10].

Fig. 2a is the schematic plot of the Fermi surface of $La_{2-2x}Sr_{1+2x}Mn_2O_7$ ($x \sim 0.59$) [13]. Contrary to the bi-layer split band structure we reported for the $x = 0.36$ and $0.38$ compounds [14], the Fermi surface of the $x \sim 0.59$ compound consists of only one hole pocket (primarily of in-plane $d_{x^2-y^2}$ states) around the zone corner, owing to its AAFM spin ordering which blocks the coupling between neighboring $Mn$-$O$ planes [13]. Figs. 2b and 2c show the near-$E_F$ band dispersion along the blue and red cuts in Fig. 2a, respectively. The variation of these bands with temperature (see Figs 2d and 2e) sheds light on the influence of stripes on the electronic excitations. Near the zone boundary, as shown in Fig. 2d, the dispersive $d_{x^2-y^2}$ band is very clear at low temperature. With increasing temperature, the dispersive $d_{x^2-y^2}$ states diminish and some localized weight continuously increases and eventually a non-dispersive/localized feature around -0.8 eV (indicated by the shaded area) becomes dominant above $T_c$. Similar behavior also occurs along the zone diagonal (see Fig. 2e).

On the other hand, the related CE stripe-like ordering observed in the $x = 0.50$ sample does not localize the electrons, as shown in Figs. 2f and 2g. In this case the sample transitions from an AFM CE stripe-ordered state at low temperature to a PM CE stripe-ordered state at intermediate temperature [12]. The difference between CE and bi-stripe on the localization of the electrons can be understood by the schematics of Figs. 1b, 1c and 1d. In these schematics we draw in the zigzag conduction paths. A long hopping path is allowed for the $x = 0.5$ sample, giving a Bloch-like dispersive band, while these paths are destroyed for the $x = 0.59$ sample due to the extra $Mn^{4+}$ ions embedded between charge stripes. This gives rise to the localized states we observe in the bi-stripe phase. Here we also note the influence of the spin order on the electronic structure is not dominant — Figs. 2f and 2g show a weak modification upon the change of magnetic order.

A slight bit of remnant dispersive $d_{x^2-y^2}$ weight persists above $T_c$ for the $x = 0.59$ sample (see Fig. 2h), which can be empirically resolved by removing a non-dispersive weight represented by an energy distribution curve (EDC) along the cut 3 in Fig. 2b. Though dispersive, these weak Bloch-like states above $T_c$ are not metallic, i.e. they do not cross $E_F$. Fig. 3 quantitatively reveals the evolution of spectral weight with increasing temperature. Figs. 3a-c show EDCs taken at representative momentum positions (EDC cuts 1, 2, and 3) in Fig. 2b. EDC 1 runs across the band bottom, EDC 2 is taken at $k_F$, and EDC 3 is in the unoccupied $k$ range. All spectra were normalized by incident photon flux only. As the temperature increases and the long-range bi-stripe structure is approached, the near-$E_F$ spectral weight diminishes and extra spectral weight grows around -0.8 eV. Fig. 3d shows "differential" EDCs, obtained by subtracting EDC cut 3 from EDC cut 2, giving an emphasis on the dispersive $d_{x^2-y^2}$ states (similar to what was done to make the plot of Fig. 2h). Here, the dispersive spectral weight continuously diminishes with increasing temperature and a gap of ~100meV opens just above $T_c$. Unlike Fig. 3b, there is no significant spectral weight built up around -0.8 eV, which suggests that the growing weight at this energy in Fig. 3b is dominated by localized weight that is similar to Fig. 2d$_4$. Near the zone center, along the red cut in Fig. 2a, a similar loss of the spectral weight has also been observed (not presented here). Here we note that at other doping levels the dispersive band is dominant in ARPES spectra, e.g. $x = 0.38$ [15], $x = 0.4$ [16,17] and the systematic change as indicated in Fig. 3c is absent. For instance, Fig. 3e plotted EDCs at $k > k_F$ taken from $x = 0.38$ samples, and there is no evident change with temperatures at ~ -0.8eV. In the $x \sim 0.59$ compound, this notable change is comparable to that of the dispersive band, which suggests a special significance of the localized weight.

By analogy to the "isosbetic point" in optical conductivity data [18], spectral weight "fixed points" can be observed in Figs 3a-c as indicated by the red arrows. In the optical conductivity experiments such isosbectic points have typically been associated with a two-component behavior with weight transfer between the components (as opposed to a continuous evolution of properties of one component). Such a two-component behavior is natural for explaining our ARPES data as well, with the two components being the dispersive state from itinerant electrons and the non-dispersive or localized state at -0.8 eV from static localized electrons.

Fig. 4 plots the variation of spectral weight of these data as a function of temperature. The red squares show the dispersive spectral weight, i.e. the integral of the weight across the entire band of Fig. 2h, while the blue squares show the near-$E_F$ portion of this data (within 100 meV of $E_F$). The black dots show the integrated weight from -1.4 eV to -0.4 eV (the isosbectic point) of EDCs in Fig. 3c, which to a good approximation depicts the localized/non-dispersive spectral weight. The plot shows that the near-$E_F$ weight and the dispersive band weight possess a distinct break at $T_c$ — they both approximately linearly decrease when approaching $T_c$ from below, and eventually the former drops to zero and the latter weakens more rapidly above $T_c$. These characteristics of the itinerant $d_{x^2-y^2}$ states are in accord with the electronic conductivity measurements [9]. On the other hand, the localized weight grows with increasing temperature, becoming dominant in the insulating region of the static bi-stripe phase.

In this system, the band width $W/2$ (~3eV) of the coherent component (gold diamonds in Fig. 4 — see supplementary Fig. S1 for the raw data) is much larger than the energy scale (-0.8 eV) of the localized component. This sets a strong constraint for theoretical models and rules out some alternative underlying mechanisms such as Mott physics and small polaronic interactions (see supplementary materials for more discussion). In fact, our data naturally lead to a picture of electronic phase separation: the dispersive states come from itinerant electrons, while the non-dispersive/localized weight arises from bi-stripes, with the ratio of the two changing with temperature.

We emphasize that in Figs. 2d and 2e the non-dispersive/localized spectral weight above $T_c$ is a signature of bi-stripes, as also observed above $T_c$ in X-ray scattering measurements (green diamonds in Fig. 4 [9]). It is then somewhat surprising that our ARPES data show a clear localized signal below $T_c$ indicating a significant population of bi-stripe states, while the X-ray scattering measurements do not show any sign of the bi-stripe signal below $T_c$. However, we note that X-ray scattering does not actually measure the presence of individual stripes, but rather measures the correlations between them, i.e. it measures the periodicity in the spacing of the stripes. Therefore, a collection of fluctuating stripes existing below $T_c$ can be invisible to X-ray scattering experiments, while still localizing the electrons which live within the stripe regions, giving rise to the localized ARPES signal. With the stripes fluctuating, other regions of the material are expected to be occasionally absent of stripes, and these regions will contribute to the dispersive portions of electronic structure. As we raise the sample temperature, the stripe-stripe correlations strengthen and the proportion of localized weight grows, until finally at $T_c$ static stripe correlations appear, the electrons become almost fully localized within the stripes, and the colossal change in conductivity appears. Theoretical proposals for fluctuating or "nematic" electronic stripes have also been put forward for other materials [5,6] and have received a great amount of interest – the confirmation of such proposals has however been lacking. The present work breaks that logjam with, to our knowledge, the first clean evidence for such fluctuating electronic stripes.

Unlike the charge stripes in cuprates, which are electrically conductive, the electron hopping is jammed along the charge stripe in the bi-stripe phase. Combined with transport and scattering measurements [9], our data indicate that these charge stripes behave like electronic valves — the fluctuating bi-stripe components (which occur below $T_c$) do not significantly impair the overall electronic conductance; however, when they become stable and form long-range patterns (above $T_c$), the electric conductivity is heavily suppressed. These properties may suggest some approaches to tune physical properties of materials by manipulating the stripe structure. Moreover, this also hints that the quantum stripes in high-$T_c$ superconductors, if they exist, could result in a significantly different behavior from their static counterpart, e.g. superconductivity.

---

1. Zaanen, J. Self-organized one dimensionality. *Science* **286**, 251-252 (1999).

2. Emery, V. J., Kivelson, S. A. & Tranquada, J. M. Stripe phase in high temperature superconductors. *Proc. Natl. Acad. Sci.* **96**, 8814-8817 (1999).

3. Cox, S., Singleton, J., McDonald, R. D., Migliori, A. & Littlewood, P. B. Sliding charge-density wave in manganites. *Nature Materials* **7**, 25-30 (2008).

4. Cheong, S.-W. & Hwang, H. Y. in *Colossal Magnetoresistance Oxides* (ed. Tokura, Y.) Ch. 7, 237–280 (Gordon and Breach, London, 2000).

5. Zaanen, J. Superconductivity: quantum stripe search. *Nature* **440**, 1118-1119 (2006).

6. Kivelson, S. A. *et al*. How to detect fluctuating stripes in the high-temperature superconductors. *Rev. Mod. Phys.* **75**, 1201-1241 (2003).

7. Beale, T. A. W. *et al*. Orbital bi-stripes in highly doped bilayer manganites. *Phys. Rev. B* **72**, 064432 (2005).

8. Luo, Z. P., Miller, D. J. & Mitchell, J. F. Electron microscopic evidence of charge-ordered bi-stripe structures in the bilayered colossal magnetoresistive manganite $La_{2-2x}Sr_{1+2x}Mn_2O_7$. *Phys. Rev. B* **71**, 014418 (2005).

9. Li, Q. A. *et al*. F. First-order metal-insulator transition in manganites: are they universal? *Phys. Rev. Lett.* **96**, 087201 (2006).

10. Zheng, H., Li, Q. A., Gray, K. E. & Mitchell, J. F. Charge and orbital ordered phases of $La_{2-2x}Sr_{1+2x}Mn_2O_7$. *Phys. Rev. B* **78**, 155103 (2008).

11. Goodenough, J. B. Theory of the role of covalence in the perovskite-type manganites [La,M(II)]$MnO_3$. *Phys. Rev.* **100**, 564-573 (1955).

12. Li, Q. A. et al. Reentrant orbital order and the true ground state of $LaSr_2Mn_2O_7$. *Phys. Rev. Lett.* **98**, 167201 (2007).

13. Sun, Z. *et al.* Electronic structure of the metallic ground state of $La_{2-2x}Sr_{1+2x}Mn_2O_7$ for x≈0.59 and comparison with x=0.36,0.38 compounds as revealed by angle-resolved photoemission. *Phys. Rev. B* **78**, 075101 (2008).

14. Sun, Z. *et al.* Quasiparticle-like peaks, kinks, and electron-phonon coupling at the (π,0) regions in the CMR oxide $La_{2-2x}Sr_{1+2x}Mn_2O_7$. *Phys. Rev. Lett.* **97**, 056401 (2006).

15. Sun, Z. *et al.* A local metallic state in globally insulating $La_{1.24}Sr_{1.76}Mn_2O_7$ well above the metal-insulator transition. *Nature Physics* **3**, 248-252 (2007).

16. Chuang, Y. –D., Gromko, A. D., Dessau, D. S., Ki,mura, T. & Tokura, Y. Fermi surface nesting and nanoscale fluctuating charge/orbital ordering in colossal magnetoresistive oxides. *Science* **292**, 1509-1513 (2001).

17. Mannella, N. *et al.* Nodal quasiparticle in pseudogapped colossal magnetoresistive manganites. *Nature* **438**, 474-478 (2005).

18. Imada, M., Fujimori, A. & Tokura, Y. et al. Metal-insulator transitions. *Rev. Mod. Phys.* **70**, 1039-1263 (1998).

**Acknowledgements** The authors thank G. Kotliar and T. Devereaux for helpful discussions. Primary support for this work was from the U.S. National Science Foundation under grant DMR 0402814. The Advanced Light Source is supported by the Director, Office of Science, Office of Basic Energy Sciences, of the U.S. Department of Energy under Contract No. DE-AC02-05CH11231. Argonne National Laboratory, a U.S. Department of Energy Office of Science Laboratory, is operated under Contract No. DE-AC02-06CH11357. The U.S. Government retains for itself, and others acting on its behalf, a paid-up nonexclusive, irrevocable worldwide license in said article to reproduce, prepare derivative works, distribute copies to the public, and perform publicly and display publicly, by or on behalf of the Government.

Correspondence and requests for materials should be addressed to D.S.D (Dessau@colorado.edu) or Z.S (SunZhe@gmail.com).

**Fig. 1 Long-range stripe structure in real space. a**, Phase diagram of $La_{2-2x}Sr_{1+2x}Mn_2O_7$ (0.48<x<0.65) **b,c**, CE ordering at x=0.5 with AFM and PM spin order, respectively. **d**, Bi-stripe ordering at x=0.6. Red arrows indicate electron hopping paths, which are broken in x=0.6 bi-stripe phase. The doping and temperature of our study is indicated by the blue arrow in **a**.

**Fig. 2 Electronic structure measurements of $La_{2-2x}Sr_{1+2x}Mn_2O_7$ (x~0.59). a**, A schematic Fermi surface plot. **b,c**, Energy vs. momentum dispersive band taken at T~20 K along the blue and red cuts in **a**. **d,e**, Stacked EDCs from the two cuts at various temperatures; **f,g**, stacked EDCs of the $d_{x^2-y^2}$ band of x=0.5 compound, along the black cut in **a** and taken at 50 K (AFM-CE state) and 180K (PM-CE state), respectively. **h**, Bloch-like dispersive band derived from Fig. 2d$_4$ by subtracting an empirical *k*-independent "background" represented by EDC cut 3 in **b**.

**Fig. 3 Temperature dependence of electronic structure. a-c** show the EDC cut 1, 2 and 3, indicated by the white dashed lines in Fig. 2b, taken at various temperatures. The red arrows indicate "fixed" points where the spectral weight is independent of temperature. **d**, Differential EDCs obtained by subtracting EDC cut 3 from EDC cut 2. **e**, EDCs taken at a position similar to cut3 from x=0.38 samples.

**Fig. 4 Temperature dependent spectral weight from ARPES and superlattice counts from x-ray scattering (ref. 9) of $La_{2-2x}Sr_{1+2x}Mn_2O_7$ (x~0.59).** Blue and red squares represent the total (entire occupied band) and the near-$E_F$ (-0.1 eV to +0.05 eV) spectral weight respectively. Black dots show the localized weight, approximated by integrating the EDCs of Fig. 3b from -1.4 eV to -0.4 eV. They are scaled independently so that the change in weight can be qualitatively estimated. The band width $W/2$ determined by supplemental Fig. S1 is plotted as gold diamonds.

Fig. 1

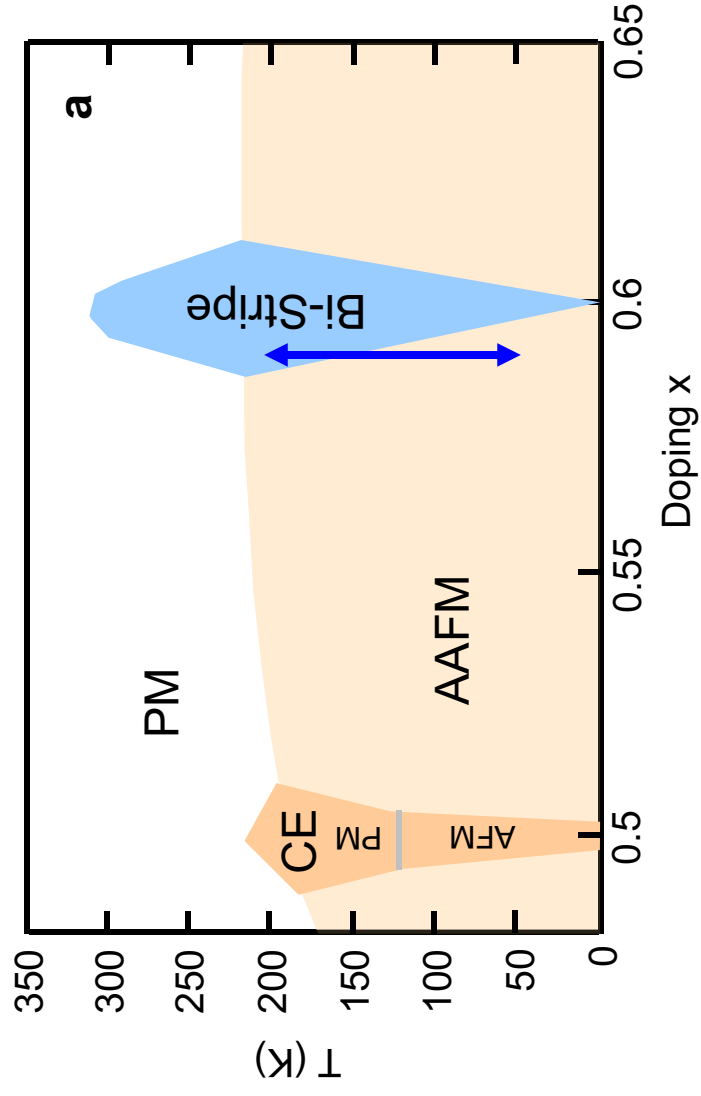
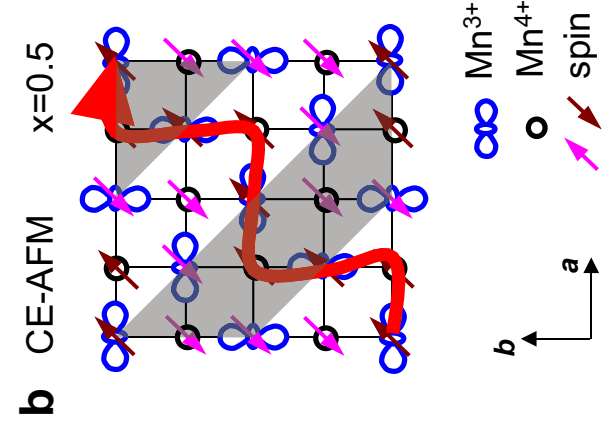

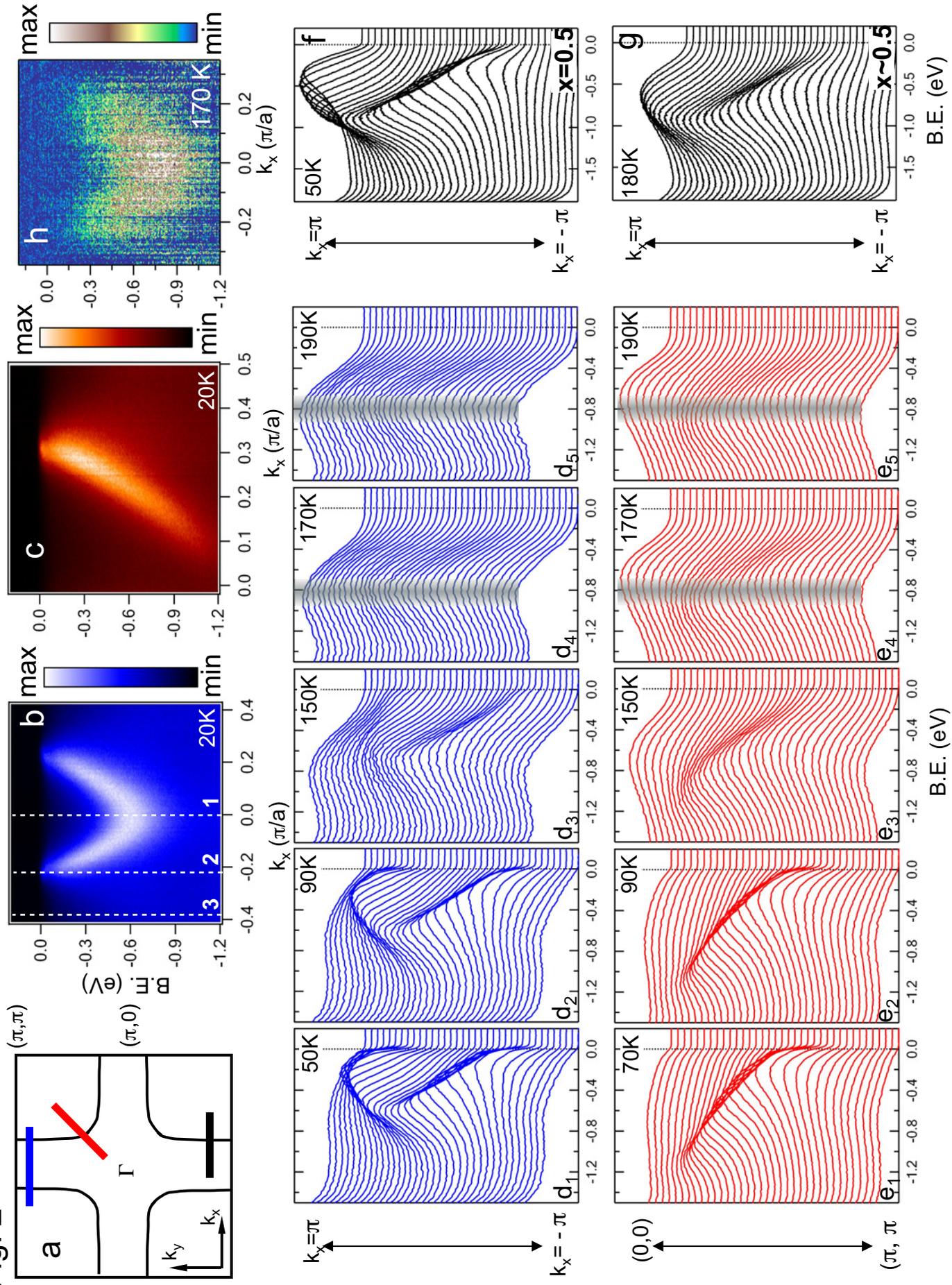

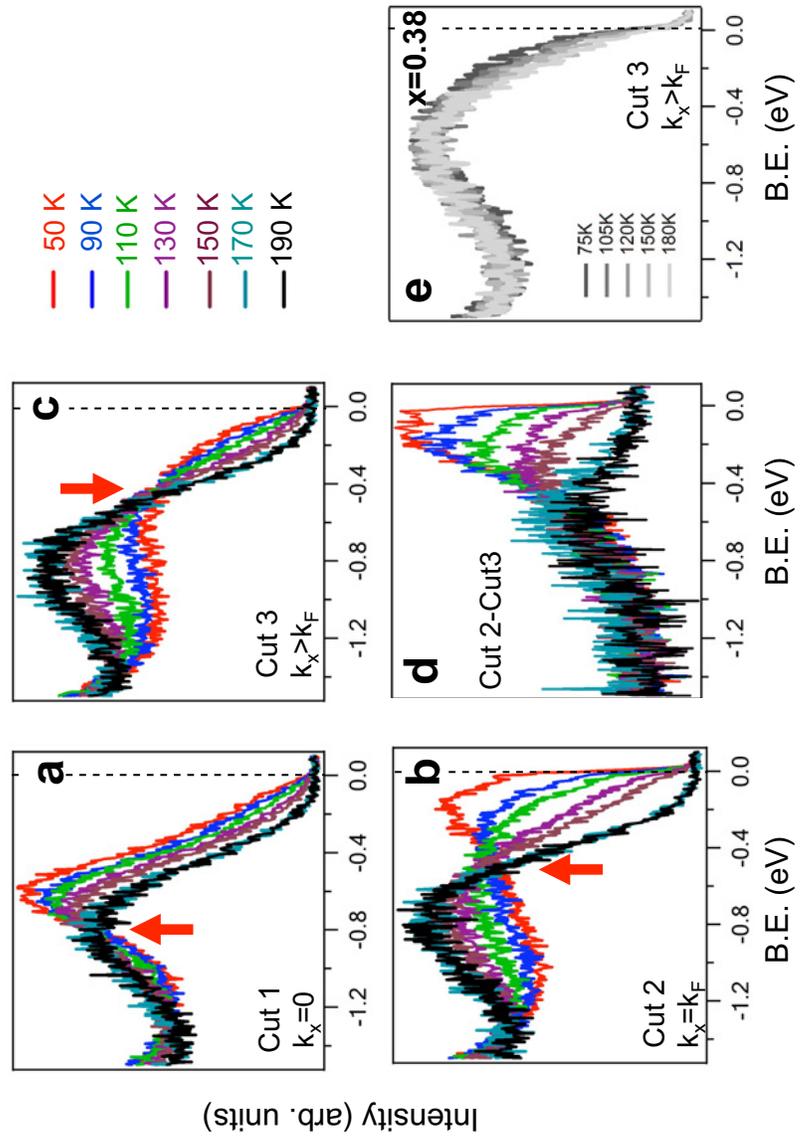

Fig. 3

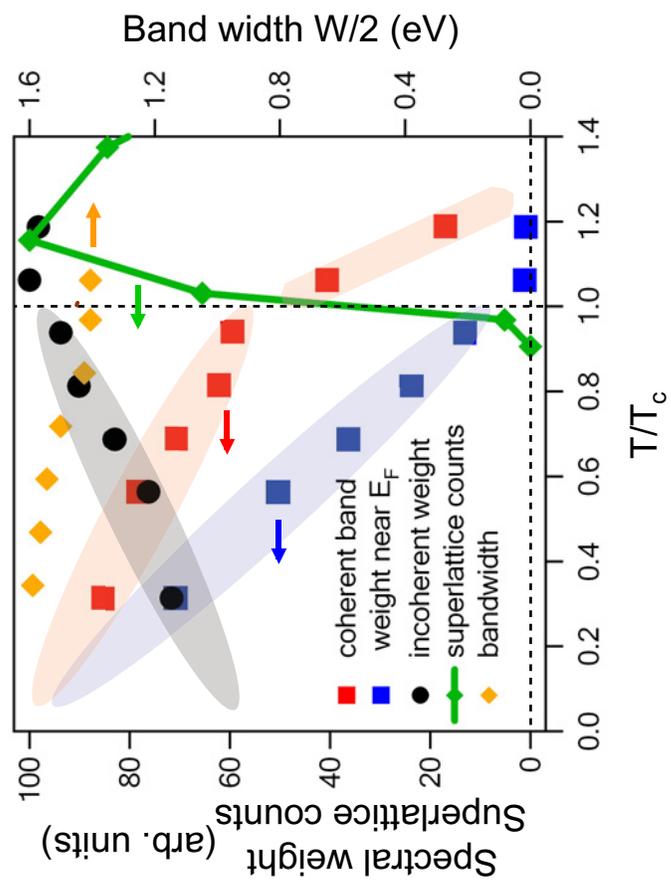

Fig. 4

# Supplementary information

## I. Experimental details.

The single crystals were grown using the traveling solvent floating zone method as described elsewhere [1]. Refined measurements suggest that the electronic properties change drastically in the vicinity of x=0.50 as well as x=0.60 [2]. In our studies, the x~0.59 samples were carefully selected and characterized as reported in ref. 3. The x=0.50 samples were characterized by transport measurements. Angle-resolved photoemission experiments were performed at Beamline 12.0.1 of the Advanced Light Source (ALS), Berkeley, using a Scienta SES100 electron analyzer under a vacuum of ~2-3×10$^{-11}$ torr. The combined instrumental energy resolution was better than 20 meV, and the momentum resolution was about 0.02 π/a.

## II. Band structure as a function of temperature.

In addition to the spectral weight transfer out of the dispersive Bloch-like states as the temperature is raised, the overall dispersion of these Bloch-like states varies slightly with temperature as well. This can best be measured by studying the zone center states which are the farthest from $E_F$. Fig. S1(a) shows the position (diamond symbols) of the $d_{x^2-y^2}$ band bottom at (2π, 2π), with a non-dispersive weight subtracted. We see that the bottom of the band of $d_{x^2-y^2}$ states moves towards the Fermi level with increasing temperature, indicating a decrease in the bandwidth W/2 (gold diamonds in Fig. 4a). This is qualitatively consistent with a double-exchange-induced reduction in hopping amplitude due to increasing spin disorder at high temperatures [4,5]. However, as we found for x=0.4 compound, the reduction in bandwidth is smaller than what would be expected for the reduction in the sample's magnetization [6] – a behavior requiring further investigation. Fig. S1(b) shows the temperature dependence of the states at the (π,0) point. This data is obtained by subtracting EDC cut 3 from EDC cut 1 of Fig 2b. All "differential" curves are stacked for comparison, and one can notice that, in addition to the loss of the spectral weight, the bottom of the dispersive $d_{x^2-y^2}$ band moves to slightly deeper binding energies with increasing temperature. The shift of the band bottom begins to occur far below $T_C$. The systematic change is summarized in Fig. S1(c).

**III. Additional discussion on underlying physics.**

A tight-binding type picture with stripe localization and no additional interactions would place the localized electronic weight at the mean of the energies of the band. Because the band is approximately half (60% holes) full, this would be an energy very near the Fermi energy, while the experiment shows that these states appear at an energy approximately -0.8 eV. Therefore, additional physics such as Coulomb interactions (Mott physics) or small polaronic interactions must be at play. Within the Mott picture the extra 0.8 eV would come from on-site Coulomb interactions, while within the small polaron picture this would be the extra binding energy that the localized electrons gain from a local lattice distortion. A critical point here is that these interactions (Mott or small polaronic) are not on their own able to drive the system insulating - they instead depend upon the stripes to do the major portion of the work. That this is true can be seen by comparing the relevant energy scales. For a doping of ~ 60%, approximately half of the band is above $E_F$, so the full bandwidth W should be ~ 3 eV. This estimate is also supported by band calculations [7]. Similarly, the upper Hubbard band is expected to be ~U/2 above $E_F$, with U/2 ~ 0.8 eV. This suggests that the U is significantly less than the bandwidth W, while for a Mott transition without the help of the stripe localization we should have a decreasing bandwidth W as $T_c$ is approached, such that U~W at the Mott transition. Similar energy scale arguments would follow for a polaron-driven transition, or a transition involving the cooperation between polaronic and Mott physics.  In contrast, our data does not show a significant reduction in the bandwidth of the dispersive states as $T_c$ is approached, (gold diamonds in Fig. 4) implying that these Bloch-like portion of the states are not feeling significant Mott or polaronic correlations. On the other hand, once the stripes have initially localized the electrons, the bandwidth of these localized electrons is driven toward zero, implying that U/W for these already-localized states is now >>1 and Mott or polaron physics can now enter the problem. Without the bi-stripe localization the Mott or polaronic physics is not able to play a significant role.

[1] Mitchell, J. M. *et al.* Charge delocalization and structural response in layered $La_{1.2}Sr_{1.8}Mn_2O_7$: enhanced distortion in the metallic regime. *Phys. Rev. B* **55**, 63 (1997).

[2] Zheng, H., Li, Q. A., Gray, K. E. & Mitchell, J. F. Charge and orbital ordered phases of $La_{2-2x}Sr_{1+2x}Mn_2O_7$. *Phys. Rev. B* **78**, 155103 (2008).

[3]  Sun, Z. *et al.* Electronic structure of the metallic ground state of $La_{2-2x}Sr_{1+2x}Mn_2O_7$ for x≈0.59 and comparison with x=0.36,0.38 compounds as revealed by angle-resolved photoemission. *Phys. Rev. B* **78**, 075101 (2008).

[4] Zener, C. Interaction between the *d*-shells in the transition metals. II. ferromagnetic compounds of manganese with perovskite structure. *Phys. Rev.* **82**, 403 (1951).

[5] Anderson, P. W. & Hasegawa, H. Considerations on double exchange. *Phys. Rev.* **100**, 675 (1955).

[6] Saitoh, T. *et al.*. Temperature-dependent pseudogaps in colossal magnetoresistive oxides. *Phys. Rev. B* **62**, 1039-1043 (2000).

[7] Sun, Z. *et al.* Electronic structure of the metallic ground state of $La_{2-2x}Sr_{1+2x}Mn_2O_7$ for x≈0.59 and comparison with x=0.36,0.38 compounds as revealed by angle-resolved photoemission. *Phys. Rev. B* **78**, 075101 (2008).

Fig. S1  **a**, EDCs of x=0.59 $La_{2-2x}Sr_{1+2x}Mn_2O_7$ taken at the band bottom ($2\pi,2\pi$) at various temperatures, with a background subtracted to make the band bottom more visible. **b**, Differential EDCs at the ($\pi,0$) point obtained by subtracting EDC cut 3 from EDC cut 1. **c**, Schematic plot of the change of the dispersion of the $d_{x^2-y^2}$ band and the weight transfer from the dispersive $d_{x^2-y^2}$ band to the non-dispersive feature (hatched area) with increasing temperature.

Supplemental figure S1

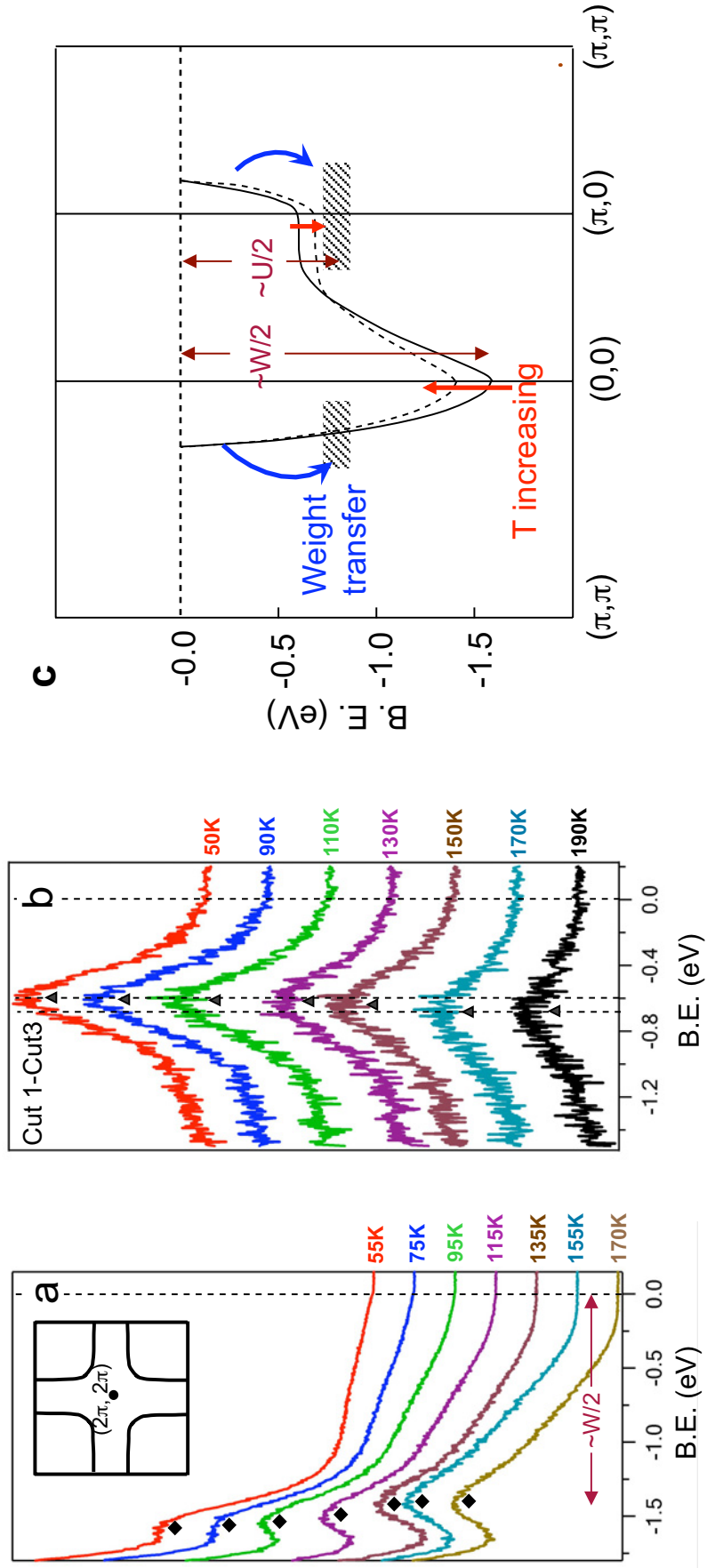